\documentclass[12pt]{iopart}
\usepackage{graphicx}
\usepackage{amssymb}
\begin{document}
\title[The dual-gain mode in CT imaging]{The dual-gain mode: a way to enhance the dynamic range of X-ray detectors}

\author{Evangelos Matsinos and Wolfgang Kaissl}

\address{Varian Medical Systems Imaging Laboratory GmbH, T\"{a}fernstrasse 7, CH-5405 Baden-D\"{a}ttwil, Switzerland}
\ead{evangelos.matsinos@varian.com and wolfgang.kaissl@varian.com}

\begin{abstract}
Varian Medical Systems has manufactured and recently put into operation a clinically-applicable solution for image-guided 
radiation therapy. Cone-beam CT imaging, one of the operation modes of the imaging unit of this device, aims at high-quality 
volumetric reconstruction. To boost the image quality, the dual-gain mode, a successful means for enhancing the dynamic range 
of the flat-panel detectors and obtaining better results in the contrast of the reconstructed image, was developed and 
successfully tested during the last few years. The important steps in the calibration of this mode involve a correction 
to the pulse widths associated with the X-ray production, the assessment of the detector signal above which nonlinear 
effects become significant and the determination of some properties of the detector pixels, namely, of dark fields, flatness 
corrections, etc. Finally, the defect-pixel map is obtained, containing dead and flickering pixels, as well as pixels with 
properties which are sufficiently `out of range'. An effect observed in the offset correction of raw images seems to 
originate in the mismatch between the measured and extrapolated values of the dark field; a correction scheme is proposed to 
take account of this effect. In the data processing, which is achieved on the basis of the files produced in the calibration 
of the mode, the high-gain signal is used whenever meaningful; otherwise, it is substituted by the low-gain signal, properly 
scaled. Defect pixels are interpolated from their good neighbours; a procedure achieving a correction in case of small defect-pixel 
clusters is outlined.
\end{abstract}

\section{Introduction}

The detailed knowledge concerning the position of the tumour and of the vital organs during the treatment has long been one 
of the most crucial and demanding problems in radiation therapy. Disregarding patient-localisation uncertainties, one may 
identify two time scales in relation to organ motion: one pertaining to daily variations (interfraction motion), the other to 
the respiration cycle (intrafraction motion). These two scales affect differently the various regions within the human anatomy.

To overcome the absence of a monitoring process, which would be practical enough for everyday application, the oncologists add 
a margin around the target in order to maximise the probability of destroying the entire tumour by the end of the treatment. 
The inevitable consequence of increasing the volume of the region which receives the largest radiation impact is the destruction 
of healthy tissue.

\begin{figure}
\begin{center}
\includegraphics [width=15.5cm] {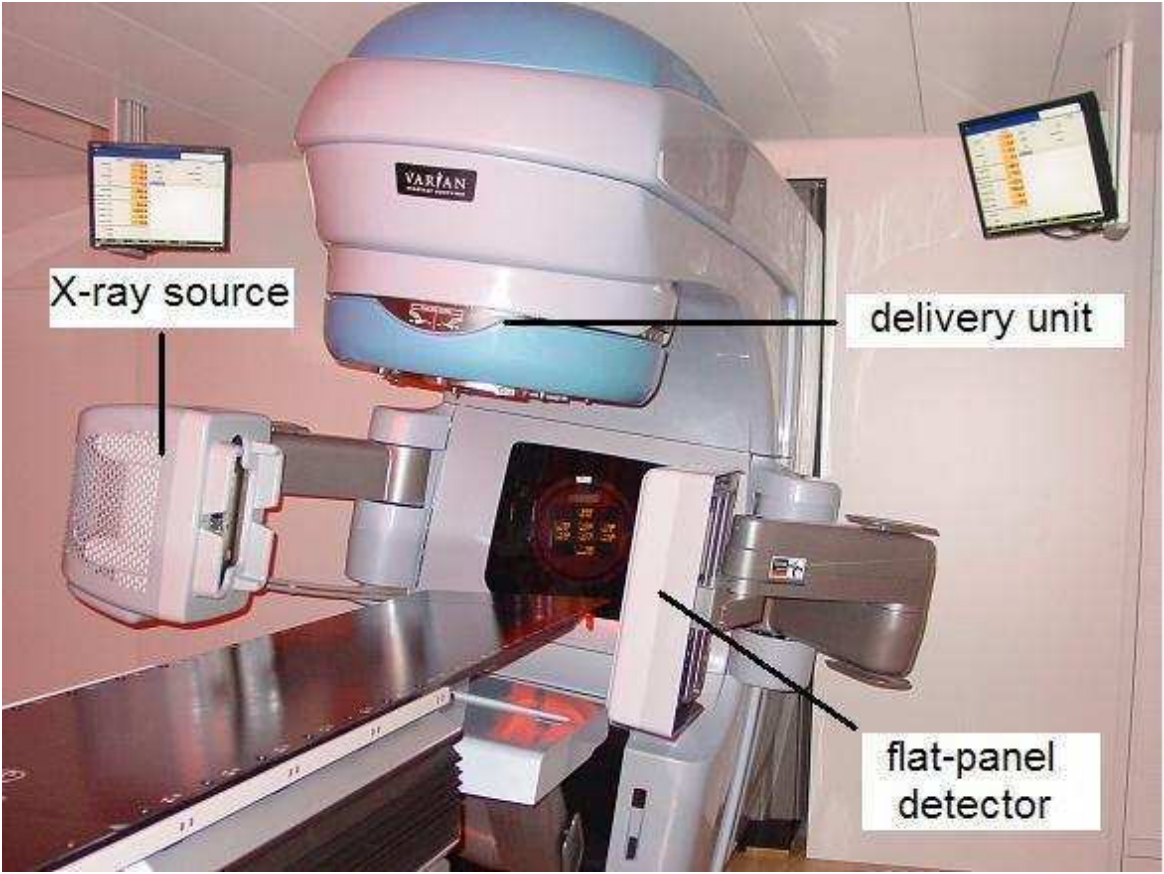}
\caption{\label{fig:OBI}The VMS Clinac accelerator equipped with imaging functionality; Hirslanden Klinik, Aarau, Switzerland.}
\end{center}
\end{figure}

Aiming at increasing the precision in radiation therapy, Varian Medical Systems, Inc.~(VMS), Palo Alto, CA, has recently deployed 
a number of new products. In autumn 2004, the VMS Clinac accelerator, equipped with imaging functionality~\footnote{The idea of 
integrating the imaging and delivery units had been proposed almost one decade ago as a means to facilitate the deposit of higher 
dose in the target and ensure better protection of the surrounding tissue; for an overview, see Jaffray \etal (2002).}, the `On-Board 
Imager' (OBI), came into operation (figure \ref{fig:OBI}). As this device is capable of both tracking and targeting tumours, 
it represents a clinically-applicable solution for image-guided radiation therapy.

The two main components of the OBI system are the low-energy (kV) X-ray tube (simply called `tube' hereupon) and the flat-panel 
detector (FPD); both are attached to the body of the device via a system of robotic arms enabling 2D (for the tube) or 3D 
(for the detector) movement. The detector which is currently in use may be operated in fluoroscopic (real-time moving images, 
to be used for position and verification) or radiographic mode (high-resolution images, to be used for diagnosis and planning).

The availability of reliable large-area FPDs at the commercial level has been a crucial factor in the development of the 
high-resolution volumetric reconstruction. The transition from traditional computed tomography (CT) to modern imaging modalities 
is well under way. In circular cone-beam CT (CBCT) imaging, the algorithm of Feldkamp \etal (1984) is a good starting point for 
further investigation; variants of this algorithm have appeared in the literature, e.g., see Grass \etal (2000) and Tang \etal (2005). 
The problem of the reconstruction in case of a helical source-detector trajectory has been studied for a number of years, e.g., 
see Fuchs \etal (2000) and the corresponding references in Grass \etal (2000). New techniques, aiming at improving the image 
quality by suppressing the noise, have surfaced with the papers of Pan (1999), Kachelrie\ss{} \etal (2001), and Pan and Yu (2003). 
Finally, renewed interest in `$180^0$ plus' reconstructions was triggered with the works of Yu and Pan (2003), and Zou \etal (2005).

To yield reliable information, the imaging unit must be properly calibrated; the calibrations may be classified as general and 
mode-specific. The general calibrations include the geometric calibration (yielding the values of important spatial parameters, 
e.g., of the source-axis and source-detector distances, as well as the corrections due to the mechanical imperfection), the $I_0$ 
calibration (yielding the open-geometry signal - the so-called `air counts' - as a function of the position on the detector), 
the norm-phantom calibration (yielding images of a norm phantom, that is, of a cylindrical object of known density, which may 
subsequently be used as reference in the reconstruction of other objects), the calibration for the beam-hardening 
correction~\footnote{Due to the energy dependence of the photon absorption length, the low-energy components of a polychromatic 
X-ray beam, when passing through matter, experience higher attenuation than the high-energy ones. This effect creates beams 
which are progressively richer (harder) in high-energy components.}, etc. The mode-specific calibrations relate to the different 
operation modes of the detector. An overview of the calibrations of the imaging unit, including some details, may be obtained 
from Matsinos (2005).

The present paper deals with the description of the most promising operation mode of the VMS FDPs, namely, the dual-gain 
mode. This mode has been developed in the recent years as a result of the need for enhancing the dynamic range of the 
detectors in order to improve image quality; the broader the dynamic range of an imaging device, the higher is the contrast 
capability in the reconstructed image. The method described here is general and may be used in systems where a dual signal, 
relating to the same object, is produced, either via successive exposures or by sophisticated readout.

\section{Materials and methods}

\subsection{Hardware components in the imaging device}

The X-ray source is the VMS model G242; it is a rotating-anode X-ray tube with maximal operation voltage of $150$ kV. The 
anode is a disc ($\varnothing$ $102$ mm) with a rhenium-tungsten face on molybdenum substrate with graphite backing. The 
tube is driven and controlled by the X-ray generator.

The VMS PaxScan 4030CB amorphous-silicon FPD is a real-time digital X-ray imaging device comprising $2048 \times 1536$ square 
elements (pixels) and spanning an approximate area of $40 \times 30$ cm$^{2}$. In order to expedite the data transfer and 
processing, the so-called half-resolution ($2 \times 2$-binning) mode is used in most of the VMS applications; thus, the detector 
is assumed to consist of $1024 \times 768$ (logical) pixels (pitch: $388$ $\mu$m). Due to the high sensitivity of the 
scintillating material (thallium-doped cesium iodide) and to the sophisticated noise-reduction techniques implemented, the 
low-dose imaging performance of this type of detector is remarkable, save for a small band neighbouring its borders~\footnote{Due 
to fluctuations in the sensitivity of the scintillator, the data within $2.91$ mm of the detector borders (the so-called 
`inactive area of the detector') are not taken into account.}. The application of this FPD in CBCT imaging has been outlined 
in Roos \etal (2004).

\subsection{The dual-gain mode}

The VMS FPDs may be operated in a number of modes; a mode is identified as a set of options pertaining to binning, gain choice, 
etc. In the dual-gain mode, the charge stored in the semiconductor is read out by using two different capacitors ($0.5$ and 
$4$ pF), leading to the so-called high- and low-gain signals, see Roos \etal (2004). In this mode, each scan (approximately 
$650$ images) produces about $2$ GB of data which has to be transferred (from the acquisition system) to the workstation for 
additional processing (correction, storage and reconstruction). The mode has been subjected to extensive and successful 
experimentation during the recent years and, at present, it represents the best option for high-quality radiography.

The dual-gain mode is an important option because it increases the dynamic range of the detector. Affecting the contrast and 
the signal-to-noise ratio, the dynamic range in an image is a key factor for high-quality CT images. The storage capacity 
of the each individual detector pixel does not impose limits to the delivered dose; the saturation level in the readout 
circuits defines (depending on the gain chosen) these limits. The optimal acquisition settings for a given exposure are those 
for which the maximal low-gain signal detected is close to the saturation level of the photodiode capacity. In the low-gain mode, 
the digital increment is coarse and restricts the contrast resolution in the high-attenuation areas of the irradiated object. To 
improve the contrast in these areas, a higher gain must be used; this is exactly where the dual-gain mode comes in. In this mode, 
each row of detector pixels is read out twice: first in low, then in high gain. One image is finally created, in which the two 
signals are interlaced.

\subsection{The calibration of the dual-gain mode}

The purpose of the calibration is to produce a set of files needed in the data processing. Before delving into the description 
of these files, we will give one definition. Henceforth, `map' implies an array of values (matrix) pertaining to the logical 
pixels of the detector; for instance, in the $2 \times 2$-binning mode, a map comprises $1024 \times 768$ elements.

The description of the calibration is facilitated if one commences with two notions which are common in imaging: the 
dark current and the flatness correction.

\subsubsection{The dark current.}

The dark current is also referred to as dark field. It is due to a number of physical processes which occur inside 
the detector irrespective of the external radiation; these processes involve effects relating to the de-trapping current, 
thermal noise, on-chip electronic noise, etc. A useful overview of these effects may be obtained from Young \etal (1998). As a 
result, the detector signal is not vanishing in the absence of external radiation. The dark-field correction is often referred 
to as offset correction of raw data.

To subtract the appropriate dark field from the data, one has to reassess its contribution frequently during the daily operation 
of the imaging unit; the dark field is sensitive to temperature variation and exposure history. To ensure that the appropriate 
dark-field values be used in the correction of the data, VMS has implemented an automatic procedure leading to the recalibration 
of the dark field if the imaging device is inactive for some (user-defined) time; additionally, the recalibration of the dark 
field is forced in case of mode switching, prior to any other calibration and shortly before acquiring scan data. To suppress 
noise in the determination of the dark field, a number of images (for example, $100$ maps) are averaged.

One important question one might pose is whether the dark field determined in the absence of radiation provides the appropriate 
offset correction in the case of exposures; this is not a trivial subject. One unique feature of the present work is that 
corrections to the measured dark fields (obtained in the absence of radiation) will be derived prior to applying the offset 
correction to the raw data.

\subsubsection{The flatness correction.}

Even when the appropriate dark field is deducted from the raw data, the resulting map is not expected to be flat. Assuming the 
absence of statistical fluctuation, there are two reasons for this deviation: the anisotropy in the radiation field and the 
variability in the response of the detector pixels to the incident radiation. We will now estimate the corrections which lead 
to a flat image at each intensity level in open-field geometry. The intensity $I$ may be thought of as the product of two 
acquisition settings, namely, of the tube current and pulse width.

The intensity-signal linearity implies that the raw signal $s_{ij}$, extracted from the pixel $ij$ ($i$ denotes 
the lateral position of the pixel on the detector and $j$ the longitudinal), is given by the formula
\begin{equation} \label{eq:IntensitySignal}
s_{ij} = a_{ij} I + d_{ij} \, ,
\end{equation}
where $a_{ij}$ and $d_{ij}$ are position-dependent quantities; the latter represents the dark field. Under this 
assumption, the offset-corrected signal in each pixel is defined as
\begin{equation} \label{eq:OffsetCorrectedSignal}
\overline{s}_{ij} = s_{ij} - d_{ij} = a_{ij} I \, .
\end{equation}

We previously gave reasons why the offset-corrected signal is not expected to be constant over the detector. To achieve the 
signal uniformity, one has to multiply the offset-corrected contents in each pixel by a flatness-correction factor which may 
be defined as
\begin{equation} \label{eq:FlatnessCorrection}
F_{ij} = \frac{ < \overline{s} > }{\overline{s}_{ij}} \, ,
\end{equation}
where the average of the $\overline{s}_{ij}$ signals over the entire area of the detector appears in the numerator on the 
right-hand side. Taking (\ref{eq:OffsetCorrectedSignal}) and (\ref{eq:FlatnessCorrection}) into account, the flatness correction 
may be written as
\begin{equation} \label{eq:FlatnessCorrectionReduced}
F_{ij} = \frac{ < a > }{a_{ij}} \, .
\end{equation}
Therefore, the factors $F_{ij}$ may be calculated from the slope parameters $a_{ij}$. If each $F_{ij}$ multiplies 
the corresponding offset-corrected signal of (\ref{eq:OffsetCorrectedSignal}), a constant value is obtained over 
the entire detector. The flatness correction is also referred to as flood-field or flat-field correction. 

Assuming that the input signals do not saturate anywhere on the detector (i.e., that the intensity-signal linearity holds), 
equation (\ref{eq:FlatnessCorrectionReduced}) explains why the flatness corrections extracted at one intensity lead to 
flat images at all intensity levels. To suppress noise in the determination of the flatness-correction maps, a number of 
images (for example, $100$ maps) are averaged.

\subsubsection{Application to the dual-gain mode.}

As two signals in each pixel are obtained in the dual-gain mode, two dark-field and two flatness-correction maps have 
to be introduced. The two dark-field values for the pixel $ij$ will be denoted as $d_{ij}^{HG}$ and $d_{ij}^{LG}$, 
where the superscripts have obvious meaning. Similarly, the two flatness corrections will be given by the formulae
\begin{equation} \label{eq:FlatnessCorrectionReducedHG}
F_{ij}^{HG} = \frac{ < a^{HG} > }{a_{ij}^{HG}} \, 
\end{equation}
and 
\begin{equation} \label{eq:FlatnessCorrectionReducedLG}
F_{ij}^{LG} = \frac{ < a^{LG} > }{a_{ij}^{LG}} \, .
\end{equation}

The fully-corrected high-gain signal may be obtained by the formula
\begin{equation} \label{eq:FullyCorrectedSignalHG}
c_{ij}^{HG} = \overline{s}_{ij}^{HG} F_{ij}^{HG} \, .
\end{equation}
Similarly,
\begin{equation} \label{eq:FullyCorrectedSignalLG}
c_{ij}^{LG} = \overline{s}_{ij}^{LG} F_{ij}^{LG} \, .
\end{equation}

Combining (\ref{eq:OffsetCorrectedSignal}), (\ref{eq:FlatnessCorrectionReducedHG}) and (\ref{eq:FullyCorrectedSignalHG}), 
we conclude that
\begin{equation} \label{eq:FullyCorrectedSignalFinalHG}
c_{ij}^{HG} = < a^{HG} > I \,
\end{equation}
and
\begin{equation} \label{eq:FullyCorrectedSignalFinalLG}
c_{ij}^{LG} = < a^{LG} > I \, .
\end{equation}
As expected, flat images have been obtained in both gains. From (\ref{eq:FullyCorrectedSignalFinalHG}) and (\ref{eq:FullyCorrectedSignalFinalLG}), 
it is deduced that the ratio of the fully-corrected signals is constant over the whole detector.
\begin{equation} \label{eq:RatioOfFullyCorrectedSignals}
R = \frac{c_{ij}^{HG}}{c_{ij}^{LG}} = \frac{ < a^{HG} > } { < a^{LG} > } \, .
\end{equation}

In the data processing, the signal $c_{ij}^{HG}$ from (\ref{eq:FullyCorrectedSignalHG}) will be used if the raw high-gain 
signal $s_{ij}^{HG}$ does not exceed a threshold value; otherwise, the low-gain 
signal $c_{ij}^{LG}$, scaled up by the ratio $R$ of (\ref{eq:RatioOfFullyCorrectedSignals}), will be used. The fully-corrected 
(scaled) low-gain signal may be expressed as
\begin{equation} \label{eq:FullyCorrectedSignalFinalScaledLG}
\tilde{c}_{ij}^{LG} = c_{ij}^{LG} R = < a^{LG} > I \frac{ < a^{HG} > } { < a^{LG} > } = a_{ij}^{LG} I \frac{a_{ij}^{HG}}{a_{ij}^{LG}}
\frac{ < a^{HG} > } {a_{ij}^{HG}} = \overline{s}_{ij}^{LG} r_{ij} F_{ij}^{HG} \, ,
\end{equation}
where $r_{ij}$ denotes the (position-dependent) high-to-low-gain slope ratio. This formula suggests that the final signal, 
corresponding to the use of the low-gain component in the data processing, may be put in the form of a product of 
the offset-corrected low-gain signal, high-to-low-gain slope ratio and high-gain flatness correction.

\subsection{The calibration files}

At the end of the calibration, the following maps are produced.
\begin{itemize}
\item The high-gain dark-field map; it is used in the offset correction of the raw high-gain data.
\item The low-gain dark-field map; it is used in the offset correction of the raw low-gain data.
\item The high-gain flatness-correction map; according to (\ref{eq:FullyCorrectedSignalHG}), it is used in the correction of 
the offset-corrected high-gain data and, according to (\ref{eq:FullyCorrectedSignalFinalScaledLG}), it is also involved in 
the correction of the low-gain data.
\item The low-gain flatness-correction map; it is used in the correction of the offset-corrected low-gain data. As previously 
explained, the low-gain flatness-correction map is simply the product of the high-to-low-gain slope-ratio and high-gain 
flatness-correction maps.
\item The threshold map; it contains the maximal reliable raw signal. The method to derive the threshold map will be given 
in detail later on.
\end{itemize}

An important point in the present work is that two additional maps will be introduced to correct for small effects in 
the two dark fields. All in all, eight maps are needed to process the data; the eighth calibration file is the map of defect 
pixels (defect-pixel map). When initialising the dual-gain-calibration workspace, the user has the option of selecting an 
already existing defect-pixel map and append to it new defect pixels; otherwise, a new map is created.

\subsection{The method}

\subsubsection{The input data.}

The input data comprises a number of dark-field images and a number of raw images; the latter are acquired at different 
intensity levels by varying the width of the pulse produced by the generator. There are two reasons why series of images 
are obtained. On one hand, this is done to suppress noise; on the other, the range of the signal values obtained 
in each pixel provides an estimate of the (signal) uncertainty which is to be used in the optimisation phase.

At this point, we are touching upon an important issue, namely, the numbers of images to be skipped and acquired at each 
intensity value. This discussion surfaces because the detector signals do not only depend on the incident radiation within 
the time window corresponding to the particular exposure, but also on the (short- and long-term) exposure history of the 
detector. The signal which is not directly associated with the dose delivered by the current pulse will be assumed to be a 
product of `lag effects'. Information on this effect may be found in Hsieh \etal (2000) and Overdick \etal (2001). The 
deep-seated traps in the semiconductor seem to be responsible for this bizarre behaviour, inducing two types of effects: 
those which affect the dark field and those which lead to the modification of the gain of each pixel.

Lag effects in the calibration have to be avoided for two reasons. First, they lead to erroneous calibration files, as 
the averages obtained do not match the characteristics of the asymptotic behaviour. Second, they artificially increase 
the uncertainties, which are subsequently to be used in the optimisation scheme, thus affecting the determination of the 
fit parameters in each pixel and complicating the statistical interpretation of the results of the linearity tests.

The data analysed in the present paper were acquired at the VMS laboratory in Baden, Switzerland, on December 15, 2005. 
A total of $200$ dark-field images were obtained. A total of $200$ exposures at each of $15$ pulse-width values~\footnote{The 
$2$ and $3$ msec pulse widths were avoided because of the larger corrections (due to the pulse shape) which are to be applied 
at the low end of the (pulse-width) range.} ($4$, $5$, $6$, $7$, $8$, $9$, $10$, $11$, $12$, $13$, $14$, $18$, $22$, $26$ 
and $30$ msec) were obtained with the tube voltage set to $125$ kV and the tube current to $16$ mA. To reduce the lag effects, 
only the last $50$ of these images at each intensity level were processed. The variation of the average high- and low-gain 
signals with the frame counter was found small, namely, below the $1 \%$ level; the variation in the original series ($200$ 
images) was around $15 \%$ for the high gain and $6$ to $11 \%$ (depending on the pulse-width value) for the low 
gain. Therefore, the exclusion of the first $150$ images at each intensity value leads to a significant reduction of the lag effects.

\subsubsection{The data analysis.}

The dark-field images were processed first. Each pixel was followed across the input files. Pixels with constant contents were 
marked defect. Pixels with null content in any of the images were also marked defect. Only a few pixels failed these two tests. 
The average signal across the images and the rms of each signal distribution (i.e., for each pixel) were evaluated and stored.

The processing of the raw images, obtained at different intensity levels, came next. The average signal across the 
input images and the rms of the distributions were evaluated and stored for each good pixel (that is, not already marked 
as defect), at each intensity level separately.

\begin{figure}
\begin{center}
\includegraphics [height=15.5cm,angle=270] {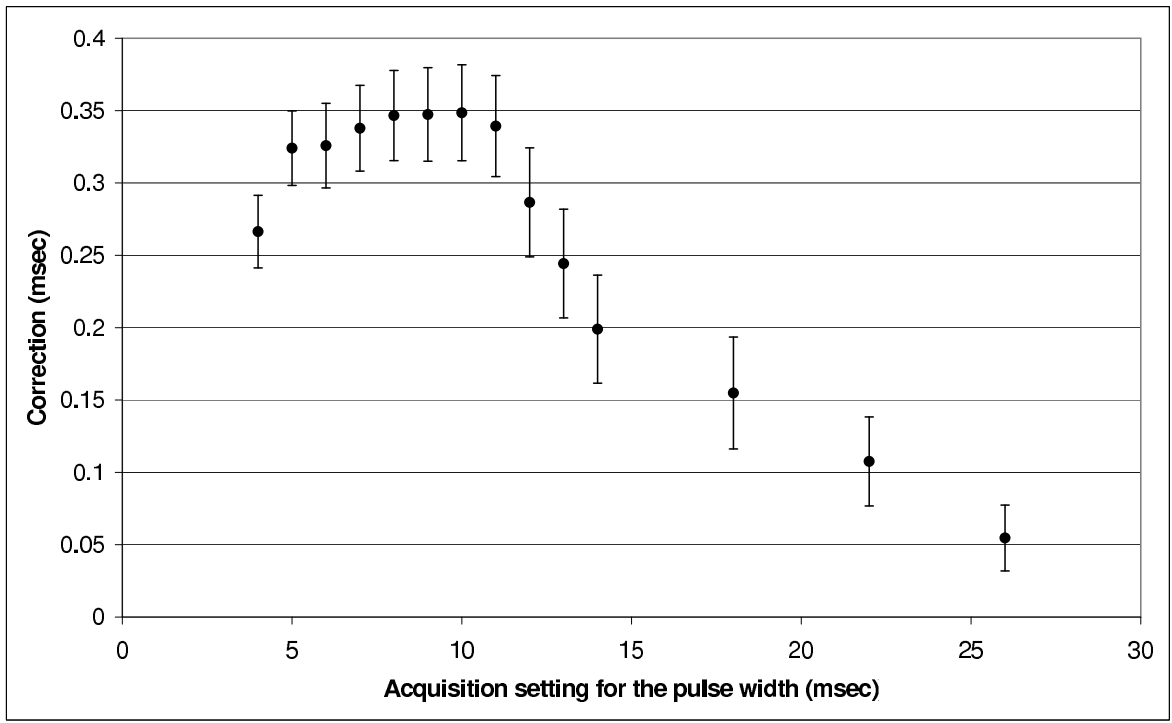}
\caption{\label{fig:CorrectionsToPulseWidths}The corrections applied to the pulse-width values in the present analysis for the various 
acquisition pulse-width settings; the uncertainties represent the rms of each distribution.}
\end{center}
\end{figure}

A correction to the pulse-width values was implemented to account for differences between the actual widths of the produced pulses 
and the nominal acquisition settings. A scheme has been put forth to derive these corrections from the low-gain signals~\footnote{In 
the future, these corrections will be estimated on the basis of the signals obtained from a normalisation chamber which is placed 
close to the tube, at the borders of the radiation field.}. For each pixel of the detector, two low-gain values were assumed `safe' 
to use: the measured dark field and the signal at the highest intensity being used in the calibration ($30$ msec). The amount by 
which each intermediate pulse-width value had to be corrected in order to bring the corresponding signal onto the straight line 
(defined by the two extreme measurements) was histogrammed, separately for the different pulse widths, and average values over the 
entire detector were obtained from these distributions. These correction factors (see figure \ref{fig:CorrectionsToPulseWidths} for 
an example) were directly applied to the high-gain signal and removed the nonlinear effects which had previously been observed.

The correction described in the previous paragraph ensures that the intensity-signal linearity is fulfilled in the 
low gain \emph{on average}; however, the absence of position-dependent effects is not guaranteed. Consequently, it does 
make sense to test the linearity hypothesis in each pixel even in the case of the low-gain data.

\begin{figure}
\begin{center}
\includegraphics [height=15.5cm,angle=270] {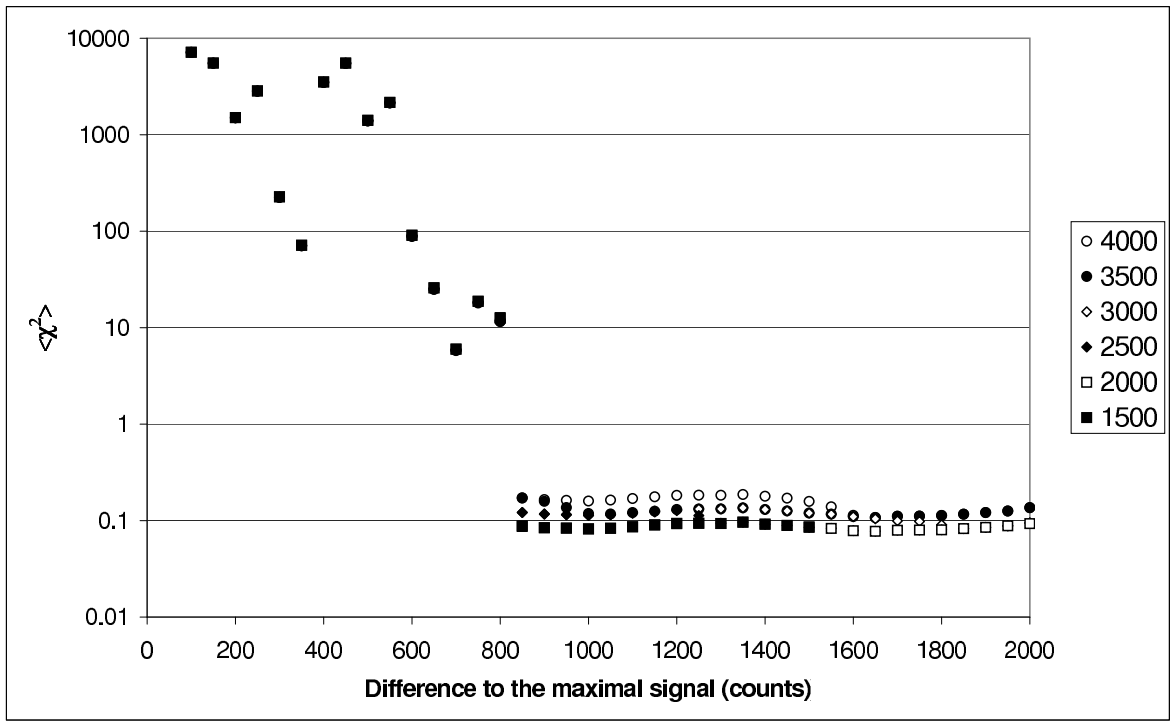}
\caption{\label{fig:AverageChi2}The average $\chi^2$ contribution as a function of the difference to the maximal signal. The series 
correspond to different choices of the parameter $d$ (see text).}
\end{center}
\end{figure}

The intensity-signal linearity is fulfilled in each pixel up to a threshold value. An absolute bound for this threshold 
corresponds to the `hard' limit in the analog-to-digital conversion, which at present utilises $14$ bits of information; 
however, nonlinearity enters the problem at smaller signals, in fact, at a fixed (that is, almost identical for all pixels) 
level below each maximal (saturated) signal $m_{ij}$. This last observation leads to two suggestions: first, differences (of 
signals) to the maximal signal (observed in the pixel chosen) should be used in this investigation and, second, a single value 
of the difference to the maximal signal (i.e., not a map) is adequate in describing the onset of the nonlinear effects over 
the entire area of the detector; in the following, this signal difference (to the corresponding $m_{ij}$) will be denoted as 
$d$. It is recommended that one be rather generous with the starting $d$ value; in the present analysis, the first $d$ value 
was $4000$ counts. The linearity hypothesis was then tested for those signals which do not exceed the $m_{ij}-d$ level. If the 
test failed, the pixel was marked defect; if it succeeded, the two parameters (slope and intercept) of the straight line, optimally 
describing the data below the $m_{ij}-d$ level, were calculated via a weighted least-squares fit. The normalised residuals for 
the signals above the $m_{ij}-d$ level were evaluated on the basis of the straight line which was obtained below the $m_{ij}-d$ 
level. Finally, the square of each normalised residual was histogrammed in bins of the difference $df$ between the chosen signal 
and the maximal signal $m_{ij}$. The position where the average $\chi^2$ contribution appeared rapidly increasing with 
decreasing $df$ identified the level at which nonlinear behaviour set in. The above procedure involves one parameter, namely, 
the distance $d$. The consistency of the scheme may be investigated by varying $d$. The average $\chi^2$ contributions as 
functions of $df$ are plotted in figure \ref{fig:AverageChi2} for different choices of $d$; it is comforting to find out that 
there is good overlap in the low-$df$ region for all $d$ choices. The conclusion from this figure is that, for the detector used, 
the nonlinear behaviour starts becoming an issue about $850$ counts below the maximal signal. To safeguard against the variability 
of the properties of different detectors (of the same type), it was decided that the threshold map contain the maximal signal 
(observed in each pixel) reduced by $1500$ counts.

A few words should be said about the linearity tests. As usual, the $\chi^2$ function is defined by the formula
\begin{displaymath} \label{eq:CostFunction}
\chi^2 = \sum_{i=1}^{n} \left( \frac{y_{i}^{m}-y_{i}^{f}}{\delta y_{i}^{m}} \right) ^2 \, ,
\end{displaymath}
where $y_{i}^{m}$ is the measured signal in one pixel, $y_{i}^{f}$ the fitted one and $\delta y_{i}^{m}$ the signal uncertainty, the 
index $i$ identifying the dose level; the sum extends over $n$ measurements. The goodness of the fit is determined on the basis of the 
p-value corresponding to the calculated $\chi^2$ value for the given degrees of freedom, that is, $n-2$, as two parameters (slope and 
intercept) are calculated (per gain type) in each pixel; the p-value of the fit represents the upper tail of the $\chi^2$ distribution 
for the given degrees of freedom. Large p-values indicate good fits, small ones point to the inefficiency of the parametric model to account 
for the input data; since a linear parametric model is used in the data description, small p-values indicate departure from linearity. In 
the present paper, statistical significance is associated with p-values below $5 \%$.

\subsubsection{Extraction of the calibration maps.}

\begin{figure}
\begin{center}
\includegraphics [height=15.5cm,angle=270] {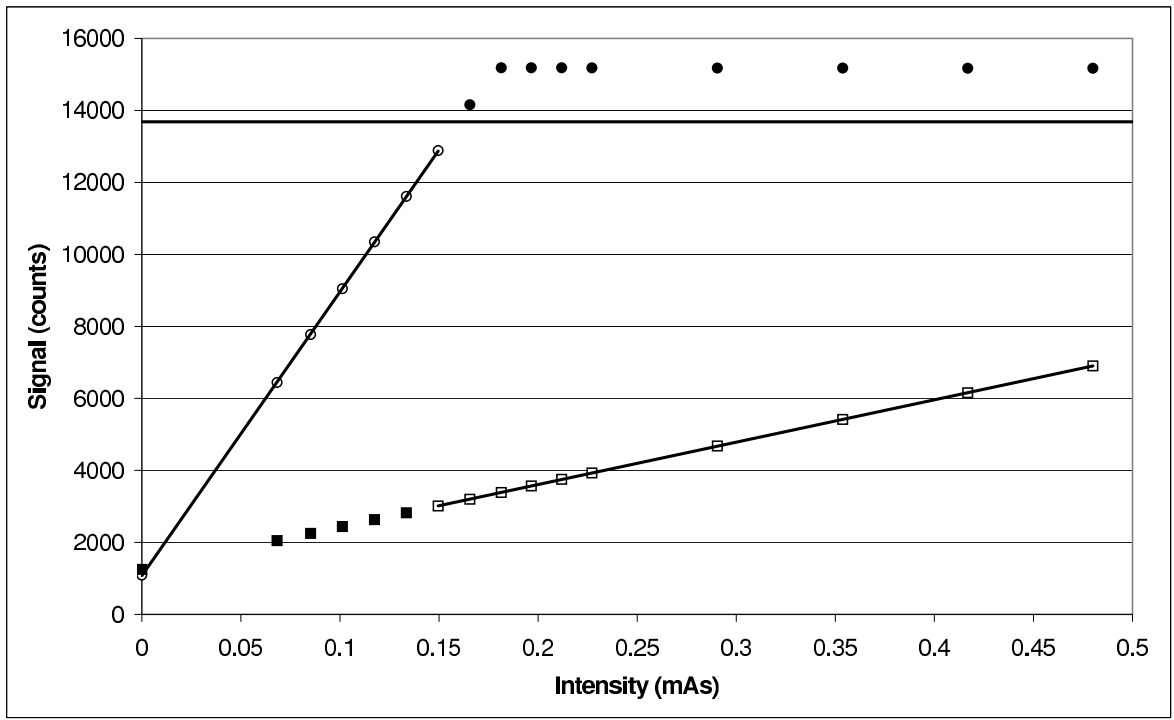}
\caption{\label{fig:SignalExample}High- and low-gain fits for one pixel chosen at random. The threshold value for this pixel corresponds 
to $13685$ counts (horizontal line). The parameters for the high gain are obtained from the signals below threshold (open circles); above 
threshold, the high-gain points (filled circles) are not used. The parameters for the low gain are obtained from the signals corresponding 
to the area where the high-gain signal saturates (open squares); the low-gain values are not used in the area where the high-gain signals 
are below threshold (filled squares). Saturation in this pixel occurs around $0.16$ mAs.}
\end{center}
\end{figure}

We now give some details concerning the extraction of the calibration maps. In each pixel, two weighted least-squares fits were performed 
(figure \ref{fig:SignalExample}). The high-gain fits were made on all the values below threshold. The low-gain fits were made from the 
highest intensity which still yielded high-gain signals below threshold onwards. This method ensures that the parameters for the 
two gains be obtained exclusively from the regions where they will subsequently (in the data processing) be used. The following 
distributions were obtained after iterating over all pixels: high-gain dark field (fitted values), low-gain dark field (fitted values), 
high-gain slope, low-gain slope and threshold.

We now come to an important point. Experience showed that the outcome of the data processing using the two measured dark fields was 
inferior to the case involving the \emph{fitted values} in the correction scheme. Taken as such, this observation implies the existence of 
a systematic difference between the measured and fitted values of the two dark fields, corroborating a previous suspicion that 
the offset obtained from a direct measurement might not correct the raw data sufficiently well. There is one additional point to 
consider. Even if the fitted values of the dark fields were stored each time that the dual-gain calibration is performed (for instance, 
on a monthly basis), a recalibration of the dark field (which is made several times during the daily operation of the imaging unit) 
would inevitably overwrite these maps. Therefore, only one set of offset files, the one obtained in the dual-gain calibration, would 
correspond to fitted values; all the subsequent files would contain measured, not fitted, values. Such a correction scheme, 
however, would be inconsistent. A solution to this problem may be obtained by storing the measured dark-field maps, as well as the 
ones corresponding to the difference between fitted and measured values. A recalibration of the dark field overwrites only the measured 
values. In the data processing, the difference of the two dark fields (assumed to be time invariant) is added to the current (measured) 
dark fields to produce the maps to be used in the offset correction of the raw data. It has to be stressed that, though the 
difference between fitted and measured dark-field values is about two orders of magnitude smaller than these values themselves, it is 
nevertheless systematic and, as such, it does affect the correction scheme. 

The high-gain flatness-correction map is obtained from the high-gain slopes via equation (\ref{eq:FlatnessCorrectionReducedHG}). 
The high-to-low-gain slope-ratio map is obtained from the ratio of the two slopes as they are extracted from the two linear fits to 
the data in each pixel. As previously said, the threshold map has been obtained from the maximal raw signal in each pixel, reduced 
by $1500$ counts.

The defect-pixel map is the last of the calibration files. As the name indicates, this map contains the coordinates of the pixels 
with unreliable signal. This failure may be due to several reasons: a pixel may be dead (constant signal), flickering (occasionally 
dead) or its properties may be sufficiently `out of range' when compared to a standard, set by the majority of the pixels (bulk of 
the data). The tests for dead pixels take place when processing the series of images being used as input in the calibration. To obtain 
a set of pixels with sufficiently close properties, we have applied cuts to the distributions of the high-gain dynamic range (threshold 
diminished by the dark-field value), low-gain dynamic range and high-to-low-gain slope ratio. The pixels which populate the tails of any 
of these three distributions were marked defect; $4 \sigma$ limits on either side of the average values were assumed. Pixels with high-gain 
flatness correction outside the range [0.5,2] were also marked defect. Although the choice of the acceptance limits (that is, identifying 
what is sufficiently `out of range') is arbitrary, the changes in our results are very small when using different cuts ($3-5 \sigma$).

The idea in the last part of the calibration is to capture those pixels which cannot be properly corrected for during a scan. This 
failure may be due to a number of reasons. For instance, as the calibration is performed at one gantry angle, it may 
be insufficient in correcting all pixels at all angles. Two scans are acquired corresponding to different acquisition pulse-width 
settings: the first scan is tailored to the high-gain signal, the second to the low-gain one. The low-gain component is not used 
in the high-gain scan; similarly, the high-gain component is not used in the low-gain scan. In either case, raw signals of about 
$7000$ to $11000$ counts are obtained in the appropriate gain type. Each pixel is followed across the images in the two scans 
separately. Pixels with constant contents are marked defect. Pixels with null content or with content below the corresponding 
dark-field value in any of the images are marked defect. Each image is then corrected using the calibration files 
obtained in the previous steps. In each image, the signal $c_{ij}$ in every good pixel is compared to the average $ < c > $ 
of its good neighbours. A dissimilarity index $\delta c_{ij}$ may be defined as
\[
\delta c_{ij} = \frac{ \mid c_{ij}- < c > \mid}{ < c > } \, .
\]
In case that $\delta c_{ij}$ exceeds $3 \%$ in any of the input images, the pixel is turned to defect. At the end of the 
analysis of the scan data, the defect-pixel map is updated (that is, it is enhanced with those pixels which were marked defect 
at this stage).

\subsubsection{The assessment of the goodness of the calibration.}

The average maps, corresponding to the pulse widths used in the calibration, are fully corrected on the basis of the calibration 
files obtained; for artefact-free image reconstructions, it is important that any deviations from uniformity on these fully-corrected 
average images be entirely due to random, not systematic, noise. An overall `noise level', defined as the mean of all rms-to-average-signal 
ratios over the entire intensity range, may serve as an indicator of the goodness of the calibration; at present, the values 
lie in the vicinity of $0.1 \%$. The pixels contained in the tails ($4 \sigma$ cut) of any of the signal distributions do not 
contribute to the calculation of this noise-level indicator; they are all turned to defect. At the end of the assessment, the 
defect-pixel map is updated. This operation completes the calibration of the dual-gain mode.

\subsubsection{Some numbers.}

The defect-pixel map was gradually enhanced as follows. The analysis of the dark-field images suggested the exclusion of $6$ pixels. 
The linearity tests failed in the case of $11$ pixels. A total of $1135$ pixels were excluded as belonging to the tails of the 
distributions of the high-gain dynamic range, low-gain dynamic range and high-to-low-gain slope ratio; the lion's share corresponds to 
the two former distributions and mostly represents three half-columns of detector pixels which stick out dramatically from the rest 
of the data. An additional $3071$ pixels were found defect during the analysis of the data of the two scans. Finally, $1180$ pixels 
failed to pass the signal-distribution cut applied during the assessment of the goodness of the calibration. Therefore, the resulting 
defect-pixel map contained $5403$ pixels, corresponding to about $0.7 \%$ of all pixels in the active area of the detector.

When fitted to a Gaussian form, the distribution of the high-gain dynamic range yielded an average of $12620$ and an rms of $120$ 
counts. The high-gain flatness correction for all good pixels was found equal to $1.003 \pm 0.052$ and the low-gain flatness 
correction to $1.002 \pm 0.049$. The high-to-low-gain slope ratio was found equal to $6.733 \pm 0.087$.

Finally, a straight line was fitted to the average fully-corrected signals, obtained in each of the average maps corresponding 
to the different intensity levels used in the calibration. The description of the data was found to be excellent and the extrapolation 
to null intensity yielded an intercept of $-2.5 \pm 1.9$ counts, compatible with the expectation value of $0$. The slope on this 
plot was found equal to $79699.3 \pm 9.7$ counts/mAs.

\subsection{The processing of dual-gain data}

The dual-gain data are processed in the following manner. In each pixel, two fully-corrected signals are created: the first corresponds 
to the high gain (\ref{eq:FullyCorrectedSignalHG}), the second to the low gain (\ref{eq:FullyCorrectedSignalFinalScaledLG}). The raw 
high-gain signal $s_{ij}$ is then compared to the threshold value $t_{ij}$ for the pixel currently processed; only if $s_{ij}$ is 
smaller than $t_{ij}$, is the high-gain content used. Finally, one map is created corresponding to half the size of the input data.

To ensure a smooth transition between the regions where the high- and low-gain signals are exclusively used, one may think of introducing 
a constant window ($w$) below threshold where both signals (with the appropriate weights) are involved in the determination of the output 
value. The weights which the two signals are to be given in the overlapping region are subject to two conditions. First, they have to be 
`continuous' functions of the input signal: the high-gain weight must equal $1$ at the $t_{ij}-w$ position and $0$ at $t_{ij}$. Second, 
the sum of the two weights must equal $1$ at all positions.

Finally, the signals in the defect pixels are not processed; their contents are obtained via interpolations using their good neighbours. 
An iterative approach has been developed to correct for defect-pixel clusters (defect pixels without any good neighbour). At each iteration 
cycle, the defect pixels with the maximal number of good neighbours are identified and corrected for; the value assigned to any such 
pixel is the average of its good neighbours. At the end of each iteration, the defect-pixel map is updated (the pixels which were 
interpolated at that step are restored as good). The process is repeated until all defect pixels have been corrected for. (Naturally, 
such a simple scheme is not expected to correct in case of large defect-pixel clusters, but then again which one would?) On the technical 
side, the correction sequence (defining which pixel is to be corrected at which step) solely depends on the supplied defect-pixel map; 
the algorithm proposed leads to unique corrected maps. To enable the fast processing of the data, the way in which the correction is to 
be performed is determined only once (at the initialisation step, prior to the data processing) and the correction sequence is stored in 
an array.

\section{Results}

To assert the improvement in the image quality when using the dual-gain signal, three reconstructions were made: the 
first two involved separately the low- and high-gain components of the dual-gain signal, whereas, in the last case, the 
two signals were combined as described in the previous section. As an example, a phantom (Catphan 500, The Phantom 
Laboratory, Salem, NY) was scanned, using the acquisition settings: $125$ kV, $80$ mA and $12$ msec. 

\begin{figure}
\begin{center}
\includegraphics [width=15.5cm] {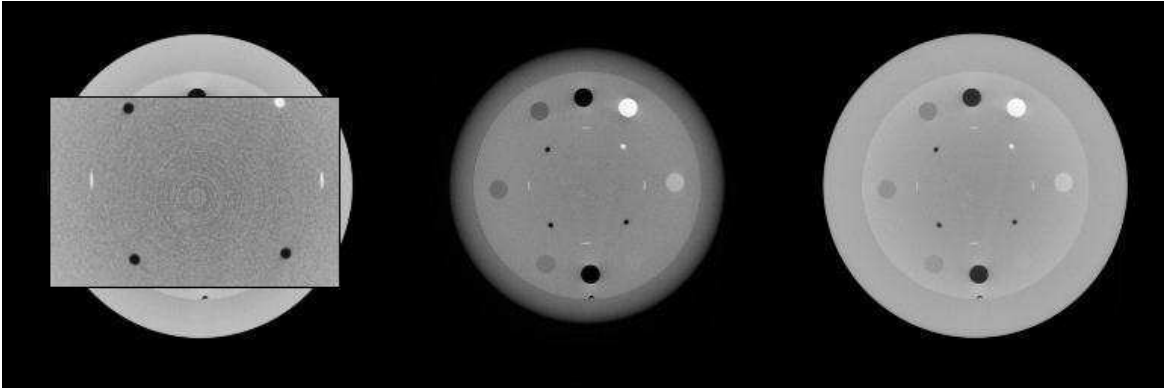}
\caption{\label{fig:Catphan}The reconstructed central slice of a phantom using the calibration and data-processing scheme described in 
the present paper (right); also shown are the results involving only the low- (left) and the high-gain (middle) components of the 
dual-gain signal. In the case of the low-gain signal, the central area of the reconstructed image has been magnified to demonstrate 
the signal degradation.}
\end{center}
\end{figure}

The results for the reconstructed central slice of the phantom (corresponding to its CTP401 module) are shown in figure 
\ref{fig:Catphan}. The reconstruction resolution is $512 \times 512$. The standard 2D weighted filtered backprojection 
algorithm was used, see Kak and Slaney (1988), combined with modified-Blackman noise filter. The input data comprised 
$675$ images obtained within one complete rotation of the gantry. The geometric calibration, see Matsinos and Kaissl (2006), 
was made prior to the data acquisition and yielded the values of the lateral displacement of the detector ($2.335$ mm) and 
source-detector distance ($1496.74$ mm); these values were used as input in the three reconstructions. In the processing of the 
dual signal, a constant window of $500$ counts below the threshold was introduced in each pixel; linear weights for the two 
gains were assumed in the overlapping window. It is evident that the reconstruction of the fully-processed dual-gain signal 
is of superior quality; when only the high-gain component is used, the borders of the phantom cannot be reconstructed (due 
to signal saturation), whereas noise effects and artefacts appear (as a result of the signal degredation in the high-attenuation 
areas of the irradiated object) in the reconstruction of the low-gain signal.

\section{Discussion and conclusions}

With image-guided radiation therapy (IGRT), tracking and targeting tumours will become simultaneous processes. Such a 
development will boost efficiency, reliability and safety in radiation therapy. To this end, Varian Medical Systems, Inc.~(VMS), 
has produced and installed the first IGRT machine, the standard VMS Clinac equipped with an imaging unit. Cone-beam CT (CBCT) 
imaging, one of the operation modes of the imaging unit, aims at high-quality volumetric reconstruction. The information on 
the position of the tumour and of the vital organs and tissue may be processed quickly, enabling modifications in the treatment 
plan on the daily basis and resulting in the delivery of higher dose in each treatment session and higher protection of the 
healthy tissue surrounding the tumour. The dynamic targeting, the simultaneous tracking and targeting tumours, is the next goal.

A number of calibrations are needed in order to operate the imaging unit properly; these calibrations ensure that the 
machine components are aligned, the mechanical distortions are small and yield important output which is to be used in 
the reconstruction of the scan data. To improve quality in CBCT imaging, VMS has developed the dual-gain mode, a successful 
means for enhancing the dynamic range of the flat-panel detectors (FPD). The broadening of the dynamic range leads to better 
contrast in the reconstructed images as meaningful data is simultaneously obtained both in the high- and low-attenuation 
areas of the irradiated object. The appropriate calibration of the mode has been achieved via an elaborate pixel-to-pixel 
analysis and has been shown to lead to improved artefact-free images.

The response of each pixel of the FPD to the variation of the incident radiation is thoroughly investigated on the basis 
of a series of images obtained at several intensity levels. At the onset of the calibration, a correction to the pulse 
widths is derived and applied systematically in the rest of the analysis. A threshold value in each pixel is identified 
as corresponding to the maximal raw signal which fulfills the intensity-signal linearity, a basic point in the data 
processing. The response parameters in the high- and low-gain-signal series are extracted from the input data and are used 
to produce the flatness corrections and the high-to-low-gain slope-ratio map. Finally, the defect-pixel map is obtained, 
containing dead and flickering pixels, as well as those pixels having properties which are sufficiently `out of range'.

One of the important corrections made in the present scheme relates to the dark field. It so happens that the dark fields 
obtained in the absence of incident radiation cannot correct the raw data sufficiently well; 
the fitted values, that is, the ones obtained via an extrapolation from nonzero intensities, seem to do a better job. A scheme 
has been proposed to ensure that the fitted values of the dark field, instead of the measured ones, be involved in the offset 
correction of raw data. The only assumption made at this point is that the (generally small) fitted-minus-measured dark-field 
difference in each pixel is time invariant, this being a reasonable claim regarding the temperature stability in the laboratory 
and the concepts involved in the electronics of the detector.

The data-processing scheme uses the maps obtained during the calibration. Concerning the combination of the high- and low-gain 
images, the high-gain signal is used whenever meaningful (not saturated); otherwise, it is substituted by the low-gain signal, 
properly scaled with the high-to-low-gain slope ratio. A constant window below the threshold value in each pixel may be introduced 
to ensure the smooth transition from the area where the high-gain signals are used to the one where they are replaced by the 
low-gain signals. In this overlapping window, both signals, with different weights, are invoked in the determination of the 
output values. Defect pixels are interpolated on the basis of the contents of their good neighbours. A simple method achieving the 
correction in case of small defect-pixel clusters has been implemented.

\begin{ack}
The dual-gain mode was developed in a time span of several years. Its potentiality in improving the quality in CT imaging was 
originally conceived by our colleagues: R E Colbeth, I Mollov, J Pavkovich, P G Roos, E J Seppi, E G Shapiro and G Zentai. During 
the development phase, important contributions were made by many; we apologise if some names which should have been 
included do not appear here. During the recent past, important contributions in this research were made by S B{\"a}hler, P Munro, 
J Richters, H Riem, R Suri, C A Tognina and G F Virshup. The data presented in this paper were acquired by H Riem.
\end{ack}

\References
\item[] Feldkamp L A, Davis L C and Kress J W 1984 Practical cone-beam algorithm {\it J. Opt. Soc. Am.} {\bf A1} 612-19
\item[] Fuchs Th, Kachelrie\ss{} M and Kalender W A 2000 Technical advances in multi-slice spiral CT {\it Eur. J. Rad.} 
{\bf 36} 69-73
\item[] Grass M, K\"ohler Th and Proksa R 2000 3D cone-beam CT reconstruction for circular trajectories {\it Phys. Med. Biol.} 
{\bf 45} 329-47
\item[] Hsieh J, Gurmen O E and King K F 2000 Investigation of a solid-state detector for advanced computed tomography 
{\it IEEE Trans. Med. Im.} {\bf 19} 930-40
\item[] Jaffray D A, Siewerdsen J H, Wong J W and Martinez A A 2002 Flat-panel cone-beam computed tomography for image-guided 
radiation therapy {\it Int. J. Radiation Oncology Biol. Phys.} {\bf 53} 1337-49
\item[] Kachelrie\ss{} M, Watzke O and Kalender W A 2001 Generalized multi-dimensional adaptive filtering for conventional and
spiral single-slice, multi-slice, and cone-beam CT {\it Med. Phys.} {\bf 28} 475-90
\item[] Kak A C and Slaney M 1988 {\it Principles of Computerized Tomographic Imaging} (New York: IEEE Press) p~86-92
\item[] Matsinos E 2005 Current status of the CBCT project at Varian Medical Systems {\it Proc. SPIE} {\bf 5745} 340-51
\item[] Matsinos E and Kaissl W 2006 The geometric calibration of cone-beam imaging and delivery systems in radiation therapy 
{\it Preprint} physics/0607018
\item[] Overdick M, Solf T and Wischmann H-A 2001 Temporal artefacts in flat dynamic X-ray detectors {\it Proc. SPIE} {\bf 4320} 
47-58
\item[] Pan X 1999 Optimal noise control in and fast reconstruction of fan-beam computed tomography image {\it Med. Phys.} 
{\bf 26} 689-97
\item[] Pan X and Yu L 2003 Image reconstruction with shift-variant filtration and its implication for noise and resolution 
properties in fan-beam computed tomography {\it Med. Phys.} {\bf 30} 590-600
\item[] Roos P G \etal 2004 Multiple-gain-ranging readout method to extend the dynamic range of amorphous silicon flat-panel 
imagers {\it Proc. SPIE} {\bf 5368} 139-49
\item[] Tang X, Hsieh J, Hagiwara A, Nilsen R A, Thibault J-B and Drapkin E 2005 A three-dimensional weighted cone beam filtered 
backprojection (CB-FBP) algorithm for image reconstruction in volumetric CT under a circular source trajectory {\it Phys. Med. Biol.} 
{\bf 50} 3889-905
\item[] Young I T, Gerbrands J J and Van Vliet L J 1998 {\it Fundamentals of Image Processing} (Delft: PH Publications) p~32-5
\item[] Yu L and Pan X 2003 Half-scan fan-beam computed tomography with improved noise and resolution properties {\it Med. Phys.} 
{\bf 30} 2629-37
\item[] Zou Y, Pan X and Sidky E Y 2005 Image reconstruction in regions-of-interest from truncated projections in a reduced 
fan-beam scan {\it Phys. Med. Biol.} {\bf 50} 13-27
\endrefs

\clearpage

\clearpage

\clearpage

\clearpage

\clearpage

\end{document}